\documentclass[useAMS,usenatbib,usegraphicx]{mn2e}
\usepackage{color}
\usepackage{multicol}

\title[Stellar mass PBHs as CDM]{Stellar mass Primordial Black Holes as Cold Dark Matter}
\author[J. L. G. Sobrinho, Augusto P.]{J. L. G. Sobrinho$^{1,2}$\thanks{E-mail: sobrinho@uma.pt }, P. Augusto$^{3}$\thanks{E-mail: sciman@med.up.pt }\\
$^{1}$Faculdade de Ci\^{e}ncias Exatas e da Engenharia, Universidade da Madeira, Campus da Penteada, 9020-105 Funchal,  Portugal\\
$^{2}$Instituto de Astrof\'{i}sica e Ci\^{e}ncias do Espa\c{c}o, Universidade de Lisboa, OAL, Tapada da Ajuda, 1349-018 Lisboa, Portugal\\
$^{3}$Faculdade de Medicina da Universidade do Porto, Al. Prof. Hern\^ani Monteiro, 4200-319, Porto, Portugal}

\begin{document}

\date{Accepted 2020 May 19. Received 2020 May 13; in original form 2020 March 12}

\pagerange{\pageref{firstpage}--\pageref{lastpage}} \pubyear{2020}

\maketitle

\label{firstpage}

\begin{abstract}

Primordial Black Holes (PBHs) might have formed in the early Universe due to the collapse of density fluctuations. PBHs may act as the sources for some of the gravitational waves recently observed. We explored the formation scenarios of PBHs of stellar mass, taking into account the possible influence of the QCD phase transition, for which we considered three different models: Crossover Model (CM), Bag Model (BM), and Lattice Fit Model (LFM). For the fluctuations, we considered a running-tilt power-law spectrum; when these cross the $\sim 10^{-9}$--$10^{-1}\mathrm{~s}$ Universe horizon  they originate  0.05--500~M$_{\odot}$ PBHs which could: i) provide a population of stellar mass PBHs similar to the ones present on the binaries associated with all known gravitational wave sources; ii) constitute a broad mass spectrum accounting for $\sim 76\%$ of all  Cold Dark Matter (CDM) in the Universe.

\end{abstract}


\section{Introduction}
\label{sec:Introduction}

The Laser Interferometer Gravitational-Wave Observatory (LIGO) identified gravitational waves emitted from the coalescence of a few binary black holes (BHs) located at distances of  $\sim 10^{2}$--$10^{3}\mathrm{~Mpc}$ 
\citep[][]{2016PhRvL.116f1102A,2017PhRvL.118v1101A,2018arXiv181112907T}. The masses of  these BHs are within the range 18--85~M$_{\odot}$, suggesting the existence of an important population of binary BHs within that mass range \citep[][]{2016ApJ...818L..22A}.
However, those masses  are larger than those of typical binary BHs formed in astrophysical scenarios at the final stage of stellar evolution of main sequence stars
\citep[e.g.][]{2016JCAP...11..036B,2018CQGra..35w5017K,2018CQGra..35f3001S,2018JCAP...09..039S,2019EPJC...79..246B}.

Considering that Primordial Black Holes (PBHs) might have formed in the early Universe as a consequence of the collapse of density fluctuations \citep*[][and references therein]{2016MNRAS.463.2348S} it is plausible to consider that, at least, a fraction of these BH binaries could be of primordial origin \citep[e.g.][]{2016PhRvL.116t1301B,2017PDU....15..142C,2019EPJC...79..246B,2020JCAP...01..031G}.

Stellar mass PBHs with less than $60$~M$_{\odot}$ have been ruled out as the prime constituent of Cold Dark Matter (CDM) under the assumption of a monochromatic  mass spectrum  \cite[i.e. all stellar mass PBHs  formed at a particular epoch, thus sharing a common mass, e.g.][]{1994ApJ...424..550D}. However, if a broad mass spectrum is allowed, then stellar mass PBHs might provide a relevant contribution to the Universe CDM \citep*[cf.][]{2016PhRvD..94h3504C}.

During the radiation-dominated epoch of the Universe ($\sim 10^{-33}\mathrm{~s}$ to $\approx 2.37\times 10^{12}\mathrm{~s}$), fluctuations of quantum origin (that were stretched to scales much larger than the cosmological horizon during inflation) can re-enter the cosmological horizon giving rise to the formation of PBHs \citep*[][]{1996PhRvD..54.6040G}, provided that their amplitude ($\delta$) is larger than a specific threshold value $\delta_{c}\simeq0.43$--0.47. 
However, during the QCD phase transition the value of $\delta_{c}$ decreases, favouring an even larger rate of PBH production \citep[][and references therein]{2016MNRAS.463.2348S}, in particular $\sim$~0.5~M$_{\odot}$ PBHs \citep*[e.g.][]{2018JCAP...08..041B,2019arXiv190402129C,2019arXiv190608217C}.

For a  given scale $k$, the horizon crossing time ($t_{k}$) is given by \citep[e.g.][]{2003PhRvD..67b4024B}
\begin{equation}
\label{horizon}
ck=a(t_{k})H(t_{k})
\end{equation}
where $a(t_{k})$ is the scale factor and $H(t_{k})$  the Hubble parameter.

The probability that a fluctuation crossing the horizon at some instant $t_{k}$ has of collapsing and forming a PBH can be written as \citep[e.g.][]{green2015}
\begin{equation}
\label{Bringmann_et_al_2001_eq5}
\beta(t_{k})=\frac{1}{\sqrt{2\pi}\sigma(t_{k})}\int_{\delta_{c}}^{\infty}
\exp\left(-\frac{\delta^{2}}{2\sigma^{2}(t_{k})}\right)d\delta
\end{equation}
where $\sigma(t_{k})$ is the mass variance at the horizon crossing time which can be written as \citep[e.g.][]{sobrinho2011}
\begin{equation}
\label{Bringmann_et_al_2001_eq2_v6}
\sigma^{2}(k)=\int_{0}^{\frac{k_{e}}{k}}x^{3}\delta_{H}^{2}(k_{c})\frac{P(kx)}{P(k_{c})}W_{TH}^{2}(x)W_{TH}^{2}\left(\frac{x}{\sqrt{3}}\right) dx
\end{equation}
where $k_{e}$ is the smallest scale generated by inflation, $k_{c}$  some suitable pivot scale, $\delta_{H}^{2}(k_{c})$  the amplitude of the density perturbation spectrum at $k_c$,  $W_{TH}$ represents the Fourier transform of the top-hat window function, and $P$  the power spectrum of the density fluctuations which, for a running-tilt power-law  spectrum (simplest version), 
is written as \citep[e.g.][]{2014PhRvD..89h3511E}
\begin{equation}
\label{power-spectrum}
P(k)=P(k_{c})\left(\frac{k}{k_{c}}\right)^{n(k)-1}
\end{equation}
with  $n(k)$, which specifies the dependence of the power spectrum on the comoving wavenumber $k$, the spectral index of the density perturbation \citep*[e.g.][]{1994PhRvD..50.4853C,2003MNRAS.342L..72B}. 
The spectral index at the pivot scale is $n(k_c)=n_0<1$ \citep[e.g.][]{2014PhRvD..89h3511E}.

 In order for a non-negligible amount of PBHs to be produced,  we must have a \emph{blue spectrum}, i.e., we must have $n(k)>1$ during some epochs \citep[e.g.][]{2003PhRvD..67b4024B}, which is consistent with the CMB
 anisotropy  \citep[e.g.][]{2014PhRvD..89h3511E,2016PhRvD..94h3504C} and, so, we write
\begin{equation}
\label{Dutching_2004_eq12a}
n(k)=n_{0}+\sum_{i\geq 1}\frac{n_{i}}{(i+1)!}\left(\ln\frac{k}{k_{c}}\right)^{i}
\end{equation}
with  the parameters $n_{1}$ and $n_{2}$ the running of the spectral index and the running of the running of the spectral index \citep[e.g.][]{2014PhRvD..89h3511E}, respectively.

Assuming that the majority of PBHs forming at a particular epoch
 have masses within the order of the horizon mass at that epoch,
then stellar mass PBHs formed when the Universe was  $\sim 10^{-5}$--$10^{-3}$~s old,
smack on the QCD epoch ($\sim 10^{-4}$~s) where we must study the  threshold $\delta_c$  in order to learn about the stellar mass PBH formation. We do so in this paper,
 following our previous work \citep[][]{2016MNRAS.463.2348S}, by using three different models for the QCD: Crossover Model (CM), Bag Model (BM), and Lattice Fit Model (LFM).

The aim of this paper, then, is to study the mass spectrum of PBHs within the extended stellar mass range 0.05--500~M$_{\odot}$, which covers all stellar mass PBHs. The paper is organized as follows: after reviewing, in Section \ref{sec:2}, some key aspects concerning the cosmological density parameter of stellar mass PBHs, in Section \ref{sec:3} we introduce our approach to the spectral index $n(k)$. 
In Section \ref{sec:4}  we present our  results on the mass spectrum of stellar mass PBHs and, in Section \ref{sec:discussion}, we conclude with a  discussion on these results. Table~\ref{tabela-inicial} sums up key  parameters that we use throughout
this paper.

\begin{table*}
\caption[]{Parameters used in this paper. References for the last column: 
[1]~\citet[][]{2014PhRvD..89h3511E}; 
[2]~\citet[][]{2018PhRvD..98c0001T}; 
[3]~\citet[][]{2016A&A...594A..20P}.
\label{tabela-inicial}
}
\center
\begin{tabular}{cp{10cm}lc}
\hline
{\bf Parameter} & \multicolumn{1}{c}{\bf Description}  & {\bf Value}  & {\bf Reference} \\
\hline
$n_{0}$ & spectral index at the pivot scale ($k_c$) & 0.9476 & [1]\\\\
$n_{1}$ & running of the spectral index & 0.001 & [1]\\\\
$n_{2}$ & running of the running of the spectral index  & 0.0226 & [1]\\\\
$k_{c}$ & pivot scale & $0.05\mathrm{~Mpc}^{-1}$ 
& [2]\\\\
$\delta_{H}^{2}(k_{c})$ & amplitude of the density perturbation spectrum at the pivot scale  ($k_c$) & $2.2\times10^{-9}$ & [3] \\\\
$\rho_{c}(t_{0})$ & critical density of the Universe at current epoch (t$_0$)  & $8.62\times 10^{-27}\mathrm{~kgm}^{-3}$ & [2] \\\\
$\Omega_{CDM}$ & Cold Dark Matter density parameter  & $0.258$ & [2] \\
\hline
\hline
\end{tabular}
\end{table*}


\section{The cosmological density of PBH}
\label{sec:2}

The PBH density parameter for PBHs formed at a given instant $t=t_{k}$ can be written as \citep[e.g.][]{1998PhRvL..80.5481N}
\begin{equation}
\label{NiemeyerJedamzik1998_eq4}
\Omega_{PBH}(t_{k})=\frac{1}{M_{H}(t_{k})}\int_{\delta_{c}}^{\infty}M_{PBH}(\delta,t_{k})P(\delta,t_{k})d\delta
\end{equation}
with $M_{H}$ the horizon mass at epoch $t_k$ and $M_{PBH}$
the PBH mass. Since
 \citep[e.g.][]{1998PhRvL..80.5481N}
\begin{equation}
P(\delta,t_{k})=\frac{1}{\sqrt{2\pi}\sigma(t_{k})}
\exp\left(-\frac{\delta^{2}}{2\sigma^{2}(t_{k})}\right),
\end{equation}
 assuming  only horizon-mass PBHs  produced at each epoch ($M_{PBH}(\delta,t_{k})=M_{H}(t_{k})$) we get, from equations (\ref{Bringmann_et_al_2001_eq5})~and~(\ref{NiemeyerJedamzik1998_eq4}) 
\begin{equation}
\label{omega_equal_beta}
\Omega_{PBH}(t_{k})= \beta (t_{k}).
\end{equation}
 Taking into account only non-evaporated PBHs  formed at $t_{k}$, we get \citep*[e.g.][]{2008ApJ...680..829R}
\begin{equation}
\label{omega_t0_original}
\Omega_{PBH}(t_{k})\left[1+z(t_{k})\right]=\Omega_{PBH}(t_{0},t_{k})\left[1+z(t_{eq})\right],
\end{equation}
where $z$ is the redshift, $t_0$ the current epoch, and $t_{eq}\approx 2.37\times10^{12}\mathrm{~s}$  the age of the Universe at the matter-radiation equality (cf. Sobrinho et al. 2016). From equation~(\ref{omega_equal_beta}) and the definition of scale factor we get
\begin{equation}
\label{omega_t0_tk}
\Omega_{PBH}(t_{0},t_{k})=\beta(t_{k})\frac{1+z(t_{k})}{1+z(t_{eq})}=\beta(t_{k})\frac{a(t_{eq})}{a(t_{k})}.
\end{equation}
The present day number density of PBHs formed at a given epoch $t_{k}$ can be written as \citep[e.g.][]{sobrinho2011}
\begin{equation}
\label{mass-spectrum}
n _{PBH}(t_{0},t_{k})=\rho_{c}(t_{0})\frac{\Omega_{PBH}(t_{0},t_{k})}{M_{H}(t_{k})}
\end{equation}
where $\rho_{c}$ is the critical density of the Universe. 
Integrating equation~(\ref{omega_t0_tk}) we get 
the global value of $\Omega_{PBH}$ evaluated at the present day (i.e. the present day value of the PBH density parameter which takes into account all non-evaporated PBHs),
\begin{equation}
\label{omega_t0}
\Omega_{PBH}(t_{0})=a(t_{eq})\int_{t_{*}}^{t'}\frac{\beta(t_{k})}{a(t_{k})}dt_{k}\:\:,
\end{equation}
where $t_{*}\sim 10^{-23}\mathrm{~s}$ \citep[PBHs formed before $t_{*}$ have already evaporated, e.g.][]{2010PhRvD..81j4019C,2014MNRAS.441.2878S} and $t'\sim 10^{5}\mathrm{~s}$ \citep[PBHs formed at $t'$ should have $\sim$~10$^{10}$~M$_{\odot}$; BH candidates with masses above such value are not known, e.g.][]{2016ApJ...818...47S}. The current mass density of such PBHs, of course, must not exceed the total mass density of the Universe. 
By a similar integration of equation~(\ref{mass-spectrum}), the present day value of the PBH number density is given by
\begin{equation}
\label{n_t0}
n _{PBH}(t_{0})=\rho_{c}(t_{0})\int_{t_{*}}^{t'}\frac{\Omega_{PBH}(t_{0},t_{k})}{M_{H}(t_{k})}dt_{k}
\end{equation}
or, if we are interested only on PBHs formed between two given instants $t_{1}$ and $t_{2}$ ($t_{*}\leq t_{1}<t_{2}\leq t' $)
\begin{equation}
\label{n_t1_t2}
n _{PBH}(t_{0})=\rho_{c}(t_{0})\int_{t_{1}}^{t_{2}}\frac{\Omega_{PBH}(t_{0},t_{k})}{M_{H}(t_{k})}dt_{k} \:\:.
\end{equation}

\section{The Spectral Index of the density perturbation}
\label{sec:3}

We consider a running-tilt power-law spectrum (equation \ref{power-spectrum}) with a spectral index given by equation (\ref{Dutching_2004_eq12a}). The observational input needed to compute the spectral index are the parameters $n_{i}$ measured at some pivot scale $k_{c}$. For
 $ i\geq 3$ the values 
 are still unknown, while the three known values are presented in Table~\ref{tabela-inicial}. Then, assuming $n_{i}=0, i\geq 5$  we write, from equation (\ref{Dutching_2004_eq12a})

\begin{eqnarray}
\label{nexpansion4}
n(k)=n_{0}+\frac{n_{1}}{2}\ln\frac{k}{k_{c}}+\frac{n_{2}}{6}\left(\ln\frac{k}{k_{c}}\right)^{2}\\\nonumber
+ \frac{n_{3}}{24}\left(\ln\frac{k}{k_{c}}\right)^{3}+\frac{n_{4}}{120}\left(\ln\frac{k}{k_{c}}\right)^{4} \: .
\end{eqnarray}
Now, the idea is to look for sets of values for $n_{3}$ and $n_{4}$ leading to relevant scenarios in terms of stellar mass PBH production, namely by seeking cases in which $n(k)$ exhibits a maximum, with $n_{max}>1$, at some point $k=k_{max}>k_{c}$. 

\begin{figure}
\centering
\includegraphics[width=80mm]{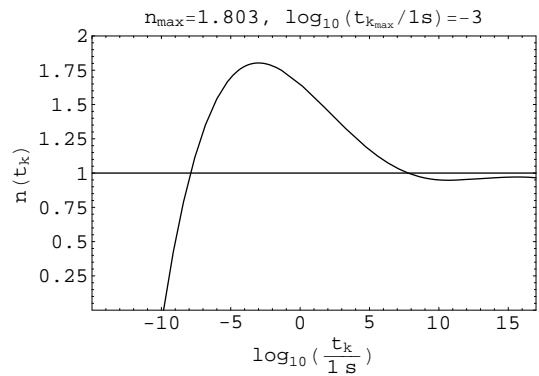} \\
\caption[]{An example of the behaviour of $n(t_{k})$ (equation~\ref{nexpansion4}) when $n_{max}=1.803$ and $t_{k_{max}}=10^{-3}\mathrm{~s}$ (the coordinates at the curve maximum). In this case we have a blue spectrum which is a requirement for PBH formation. From equations~(\ref{n3}) and (\ref{n4}) we derive $n_{3}=0.0099$ and $n_{4}=-0.0033$.
\label{fig1}}
\end{figure}

Equation (\ref{horizon}) relates a given scale $k$ with the instant $t_{k}$ and, so, we here refer to $k_{max}$ or to $t_{k_{max}}$ with the same meaning. Using $X=\ln(k_{max}/k_{c})$ we write, following equation~(\ref{nexpansion4}),
\begin{equation}
\label{nmax}
n_{max}=n_{0}+\frac{n_{1}}{2}X+\frac{n_{2}}{6}X^2+\frac{n_{3}}{24}X^3+\frac{n_{4}}{120}X^4
\end{equation}
and, by definition,

\begin{equation}
\label{dnmax}
\frac{dn}{dX}(k_{max})=
\frac{n_{1}}{2}+\frac{n_{2}}{3}X+\frac{n_{3}}{8}X^2+\frac{n_{4}}{30}X^3=0 \:\: .
\end{equation}
Solving equations (\ref{nmax}) and (\ref{dnmax})  we get
\begin{equation}
\label{n3}
n_{3}=-\frac{4\left(24n_{0}-24n_{max}+9n_{1}X+2n_{2}X^2\right)}{X^3}\:\:\: {\rm and}
\end{equation}
\begin{equation}
\label{n4}
n_{4}=\frac{20\left(18n_{0}-18n_{max}+6n_{1}X+n_{2}X^2\right)}{X^4} \:\:
,
\end{equation}
which allows us to relate ($n_{3}$,$n_{4}$) with the more meaningful quantities ($n_{max}$,$t_{k_{max}}$).
Hence, for a given pair of values ($n_{max}$,$t_{k_{max}}$) we can  determine, with the help of equations (\ref{n3}) and  (\ref{n4}), 
and the $n_0$, $n_1$, $n_2$ values of Table~\ref{tabela-inicial},
the corresponding values of $n_{3}$ and $n_{4}$ and, consequently, the curve for the spectral index $n(k)$ as given by equation (\ref{nexpansion4}).
An example in a particular case showing a blue spectrum is presented in 
Figure~\ref{fig1}.

\section{The mass spectrum of  stellar mass PBH}
\label{sec:4}

We first determine the fraction of the Universe going into stellar mass PBHs at a given epoch $t_{k}$ (cf. equation \ref{Bringmann_et_al_2001_eq5}) using the three different models of \citet[][]{2016MNRAS.463.2348S}:
i) Crossover Model (CM); ii)  Bag Model (BM); iii) Lattice Fit Model (LFM).

\subsection {Crossover Model (CM)} 
\label{sec:Crossover Model (CM)}

Equation (\ref{Bringmann_et_al_2001_eq5}) must now be written as 
\begin{equation}
\label{beta_crossover0}
\beta_{CM}(t_{k})=\frac{1}{\sqrt{2\pi}\sigma(t_{k})}\int_{\delta_{c1}}^{\delta_{c}}\exp\left(-\frac{\delta^{2}}{2\sigma^{2}(t_{k})}\right)d\delta + \beta_{R}(t_{k})
\end{equation}
where $\beta_{R}(t_{k})$ is $\beta(t_k)$ (equation~\ref{Bringmann_et_al_2001_eq5}), now seen as the \emph{contribution from radiation} (since the first term represents the contribution from the CM), while
 $\delta_{c1}<\delta_c$ is the  threshold for PBH formation, valid during the QCD phase transition in the case of a CM (cf. Sobrinho et al. 2016). For epochs sufficiently away from the transition,  $\delta_{c1}\approx \delta_{c}$ and equation (\ref{Bringmann_et_al_2001_eq5}) remains valid.

\subsection {Bag Model (BM)} 
\label{sec:Bag Model (BM)}

Now equation (\ref{Bringmann_et_al_2001_eq5}) is valid only up to some instant after which there is an additional window $[\delta_{c1},\delta_{c2}]<\delta_c$ allowing PBH formation (Sobrinho et al. 2016), written as
\begin{equation}
\label{beta_dc1_dc2}
\beta_{BM}(t_{k})=\frac{1}{\sqrt{2\pi}\sigma(t_{k})}\int_{\delta_{c1}}^{\delta_{c2}}\exp\left(-\frac{\delta^{2}}{2\sigma^{2}(t_{k})}\right)d\delta
+\beta_{R}(t_{k}).
\end{equation}
Eventually,  $\delta_{c2}$ reaches $\delta_{c}$ and we 
recover equation (\ref{beta_crossover0}).

\subsection {Lattice Fit Model (LFM)} 
\label{sec:Lattice Fit Model (LFM)}

We now consider yet another  window $[\delta_{cA},\delta_{c}]$ allowing PBH formation (cf. Sobrinho et al. 2016), writing equation (\ref{Bringmann_et_al_2001_eq5}) as
\begin{equation}
\label{beta_lattice_fit_dcA}
\beta_{LFM1}(t_{k})=\frac{1}{\sqrt{2\pi}\sigma(t_{k})}\int_{\delta_{cA}}^{\delta_{c}}\exp\left(-\frac{\delta^{2}}{2\sigma^{2}(t_{k})}\right)d\delta
+\beta_{R}(t_{k}).
\end{equation}
Over a brief interval we might also have to consider  the window $[\delta_{c1},\delta_{c2}]<\delta_{cA}$ where
\begin{eqnarray}
\label{beta_dcA_dc1_dc2}
\beta_{LFM2}(t_{k})=\frac{1}{\sqrt{2\pi}\sigma(t_{k})}\int_{\delta_{c1}}^{\delta_{c2}}\exp\left(-\frac{\delta^{2}}{2\sigma^{2}(t_{k})}\right)d\delta + \nonumber\\
+\frac{1}{\sqrt{2\pi}\sigma(t_{k})}\int_{\delta_{cA}}^{\delta_{c}}\exp\left(-\frac{\delta^{2}}{2\sigma^{2}(t_{k})}\right)d\delta
+\beta_{R}(t_{k}).
\end{eqnarray}
Again, at some point $\delta_{c2}$ reaches $\delta_{cA}$ and we return to equation (\ref{beta_crossover0}).

\subsection{Stellar mass PBHs formation (three models)}

Following equation~(\ref{nexpansion4}) (cf. Figure 1), by fixing
 $t_{k_{max}}$ we  determine the allowed range of values for $n_{max}$, i.e., the range of values of $n_{max}$ for which $\beta(t_{k})$ does not exceed the observational constraints, which will depend on the model  adopted for the QCD (equations~\ref{beta_crossover0}--\ref{beta_dcA_dc1_dc2}). 
Within the stellar mass range 0.05--500~M$_{\odot}$ these observational constraints are mainly obtained from gravitational lensing surveys, data from gravitational waves due to binary coalescences, and CMB anisotropies measured by Planck, with the maximum value allowed for $\beta(t_{k})$ on the range $[10^{-11},10^{-8}]$ \citep[for more details see][]{2020arXiv200212778C}.

\begin{figure}
\centering
\includegraphics[width=80mm]{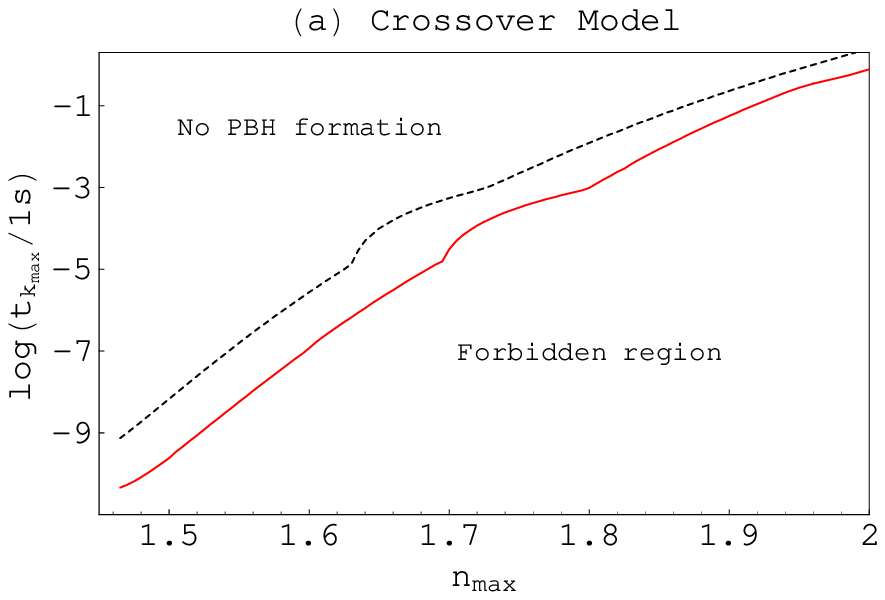} \\
\includegraphics[width=80mm]{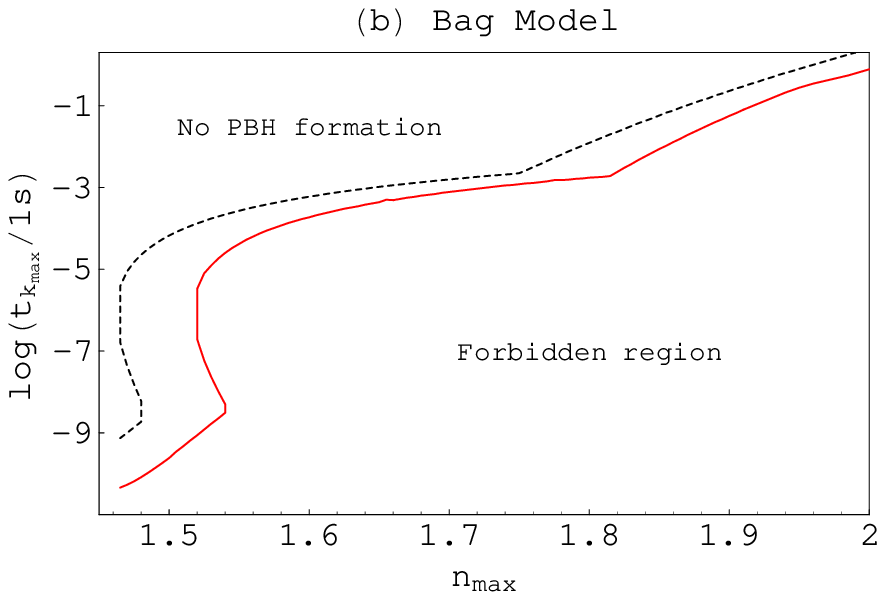} \\
\includegraphics[width=80mm]{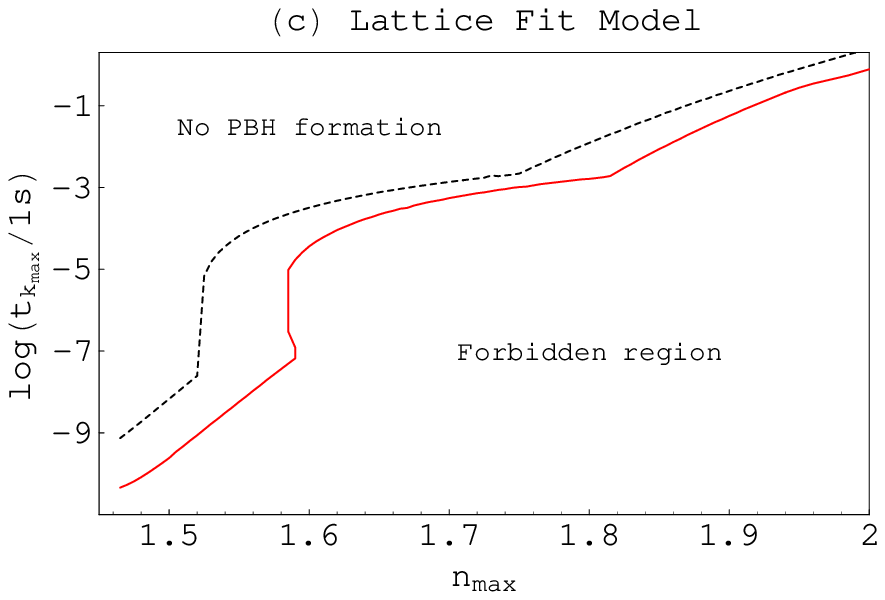} \\
\caption[]{The curve in the $(n_{max},\log(t_{k_{max}}/1\mathrm{~s}))$ plane indicating which parameter values lead to stellar mass PBH formation, according to each of the three QCD models (CM, BM, LFM). Below the solid curve, PBH formation is not allowed since it would violate the observational constraints \citep[][]{2020arXiv200212778C}. Above the dashed line, PBH formation is allowed although in negligible numbers (less than one PBH within the observable Universe). The region of interest, as regards PBH formation, is the one located between the two lines, with the most favourable situations on the solid curve: the number density of PBHs decreases as one moves from the solid line towards the dashed line. 
\label{fig2}}
\end{figure}

We found, numerically \cite[with][]{wolfram}, that in order to fully cover the extended stellar mass range 0.05--500~M$_{\odot}$ we should consider $1.4\leq n_{max}\leq 2.0$ and $10^{-9}\mathrm{~s}\leq t_{k_{max}}\leq 1$~s.
In Figure~\ref{fig2} we thus show the region on the $(n_{max},\log(t_{k_{max}}/1\mathrm{~s}))$ plane where stellar mass PBH formation is possible, between:
i) the  `forbidden region'  where the amount of formed stellar mass PBH would violate the observational constraints; ii) `No PBH formation', actually meaning that this is negligible (less than one stellar mass PBH within the entire observable Universe -- see Section~\ref{sec:5}). 

\begin{figure}
\centering
\includegraphics[width=65mm]{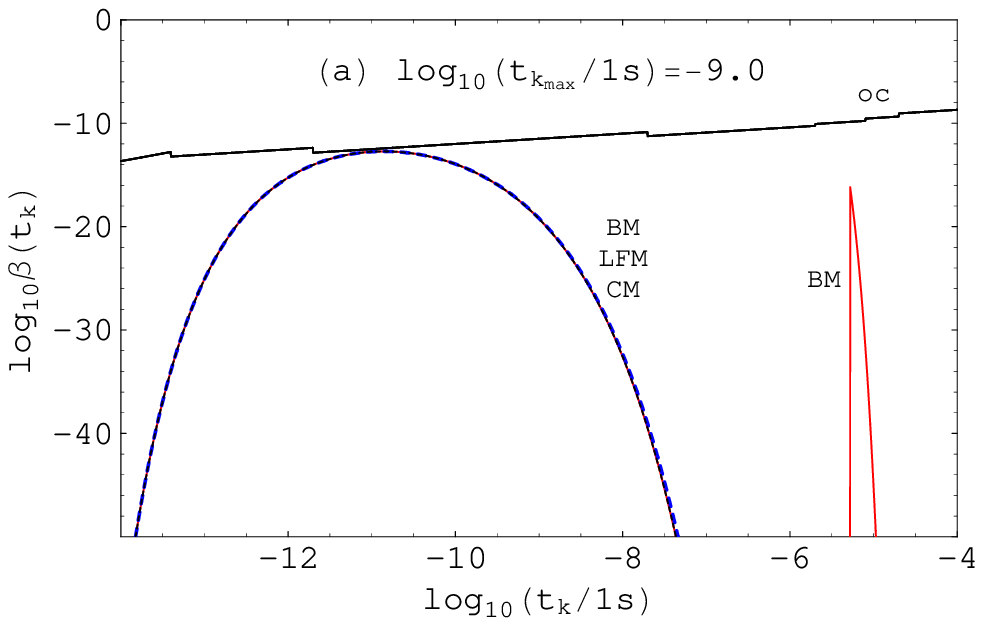} \\
\includegraphics[width=65mm]{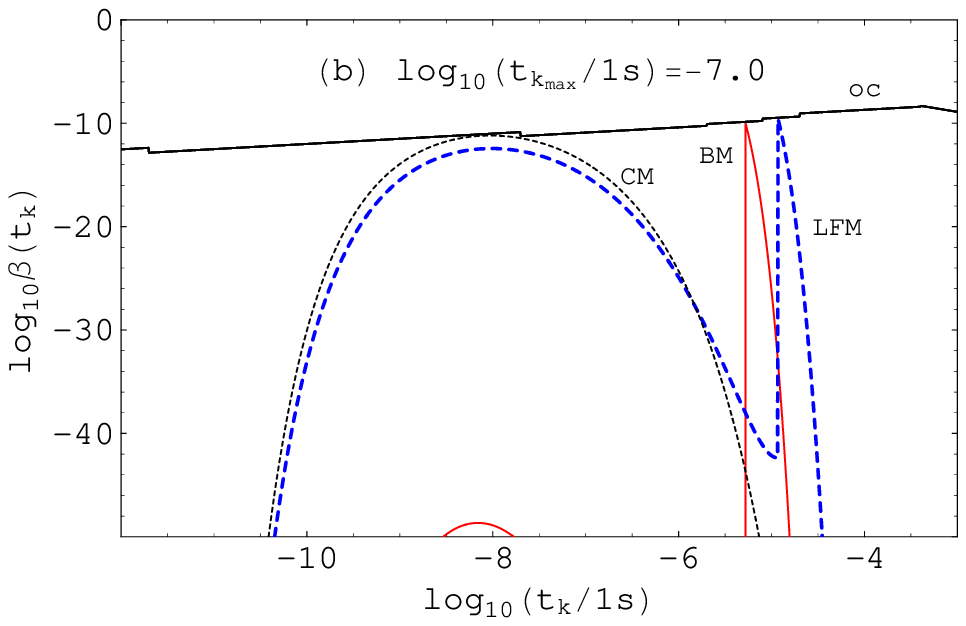} \\
\includegraphics[width=65mm]{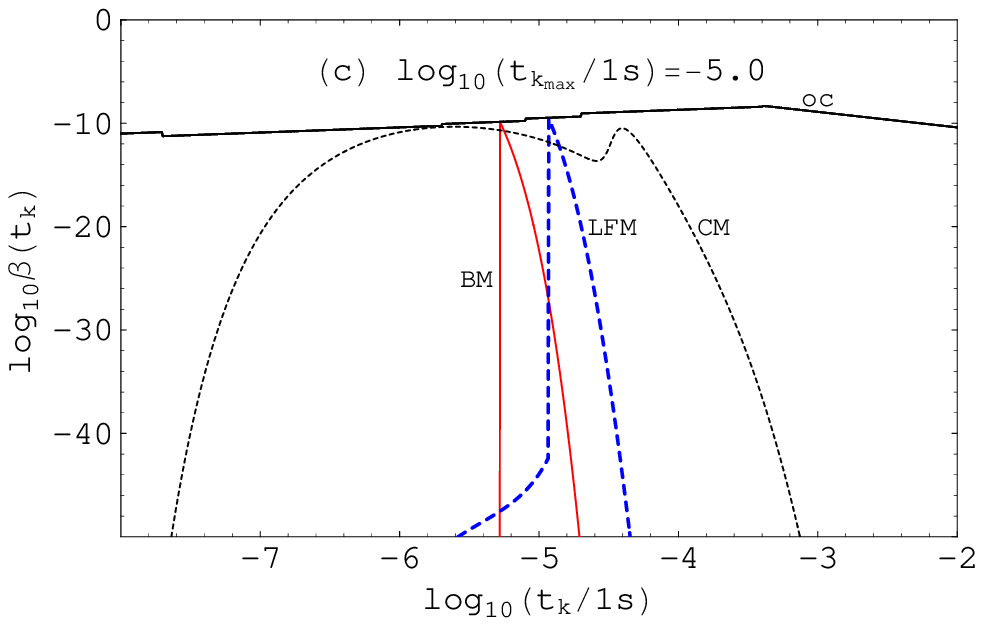} \\
\includegraphics[width=65mm]{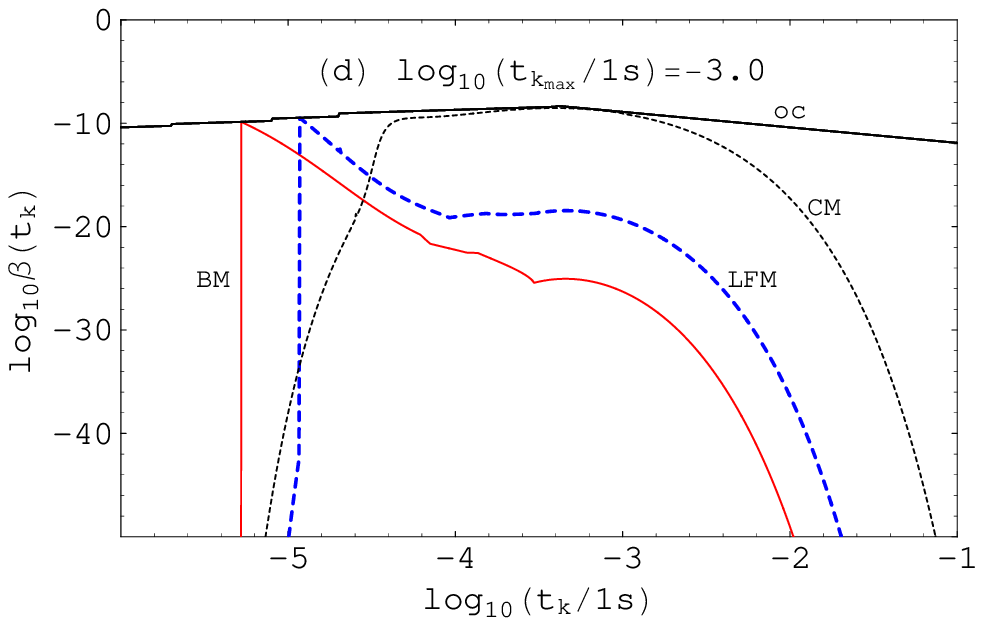} \\
\includegraphics[width=65mm]{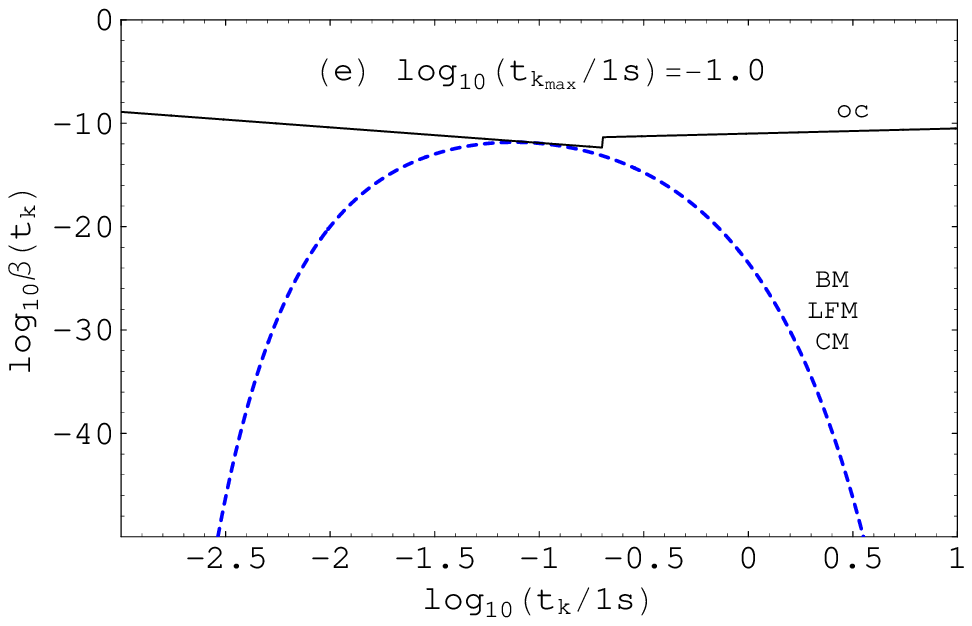} \\
\caption[]{The fraction of the Universe going into stellar mass PBHs when: 
(a) $t_{k_{max}}=10^{-9}\mathrm{~s}$; (b) $t_{k_{max}}=10^{-7}\mathrm{~s}$;
(c) $t_{k_{max}}=10^{-5}\mathrm{~s}$; (d) $t_{k_{max}}=10^{-3}\mathrm{~s}$, and (e) $t_{k_{max}}=10^{-1}\mathrm{~s}$. Each curve is labeled according to the corresponding QCD model (CM, dotted; BM, continuous; LFM, dashed). The curve labeled `oc' in the top of each figure represents the observational constraints. 
In (e) a dominating `radiation peak' is seen, centred near $t_{k_{max}}$ (for all models) while a secondary `QCD peak' arises for cases like the ones seen in (b), with the centre-of-peak dependent on the model.  Note that the situation exemplified in (d), for the CM, is the one of Figure~\ref{fig1}.
\label{fig3}}
\end{figure}

For a given value of  $t_{k_{max}}$ the fraction of the Universe going into PBHs, $\beta (t_k)$, will be maximum if the corresponding value of $n_{max}$ is the one located over the solid curve in Figure~\ref{fig2}.
Results for a selection of cases in such conditions are given in Figure~\ref{fig3} and Table~\ref{parametros-casos-estudados}.

As seen in Sections \ref{sec:Crossover Model (CM)}--\ref{sec:Lattice Fit Model (LFM)},
for epochs sufficiently away from the influence of the QCD, the dominant term in equations (\ref{beta_crossover0}) to (\ref{beta_dcA_dc1_dc2}) is $\beta_{R}(t_{k})$, and a \emph{radiation peak} is seen around $t_{k_{max}}$ (e.g.~Figure~\ref{fig3}e).
On the other hand, at epochs close to the QCD epoch, a \emph{QCD peak} 
 shows up, with the $t_{k_{max}}$ location dependent on the model (e.g. Figure~\ref{fig3}b).

When $t_{k_{max}}=10^{-9}\mathrm{~s}$ we may consider $n_{max}=1.523$ for all the three QCD models in order to maximize the number of PBHs without violating the observational constraints (Table~2). The $\beta(t_{k})$ curve for this case is shown in Figure~\ref{fig3}a. Notice that for both the CM and LFM we have the same $\beta(t_{k})$ curve (left portion of the curve in Figure \ref{fig3}a, which corresponds to a radiation peak). This happens because we are considering fluctuations that crossed the horizon sufficiently  before the QCD epoch. If we consider a BM instead, then we cannot neglect the contribution from the QCD (cf. equation \ref{beta_dc1_dc2}) and as a result we have, in addition, a QCD peak (although not quite as high as the radiation peak).

In Figure \ref{fig3}b we show the curve for $\beta(t_{k})$ when $t_{k_{max}}~=~10^{-7}\mathrm{~s}$ and with $n_{max}$  assuming the values 1.599 (CM), 1.524 (BM), and 1.593 (LFM) -- Table~2. Notice that we are considering, for each QCD model, different values of $n_{max}$ in order to reach the maximum production of PBHs allowed for each case at the considered epoch. Although the CM curve still consists entirely of a radiation peak, now the BM curve is fully dominated by the QCD peak. As for the LFM curve we have the presence ot the two peaks (namely, a radiation peak on the left and a QCD peak on the right).

Moving to Figure \ref{fig3}c we show the curves for $\beta(t_{k})$ when $t_{k_{max}}=10^{-5}\mathrm{~s}$ with $n_{max}$ assuming the values  1.686 (CM), 1.530 (BM), and 1.587 (LFM) -- Table~2. In terms of the CM we now get a radiation peak as well as a QCD peak, the latter just emerging on the right side of the curve, while the BM and LFM curves are completely dominated by their sharp QCD peaks. 

In the case $t_{k_{max}}=10^{-3}\mathrm{~s}$ (Figure~\ref{fig3}d) $n_{max}$ assumes the values 1.803 (CM), 1.732 (BM), and 1.752 (LFM) -- Table~2. In the case of a CM (see also Figure~1) we get a radiation peak and a QCD peak which join (the latter more on the left), forming some sort of plateau in the $\beta(t_{k})$ curve.
The BM and LFM are still dominated by the QCD peak, which now appears on the left, although the radiation peak is more obvious. 

Finally, in Figure \ref{fig3}e we show the curve for $\beta(t_{k})$ when $t_{k_{max}}=10^{-1}\mathrm{~s}$ and $n_{max}=1.920$. In this case we are dealing with fluctuations that crossed the horizon sufficiently after the QCD epoch and, so, all the three QCD models share the same curve (and, hence, the same value for $n_{max}$ -- Table~2) which is characterized by a radiation peak.

\begin{table*}
\caption[]{Results for a selection of studied cases which imply maximum stellar mass PBH formation (on the solid lines of Figure~2): {\bf (1)}  the instant when the fluctuation crosses the Universe horizon; {\bf (2)}  the maximum value attained by the spectral index (retrieved from Figure~\ref{fig2}); {\bf (3)} the corresponding QCD model (CM, BM or LFM); {\bf (4)} parameter $n_{3}$, from equation~(\ref{n3}); {\bf (5)} parameter $n_{4}$, from equation~(\ref{n4}); {\bf (6)} related Figure(s). The highlighted example is the one presented in Figure~1.
\label{parametros-casos-estudados}
}
\center
\begin{tabular}{cccccc}
\hline
{\bf (1)} & {\bf (2)}  & {\bf (3)}  & {\bf (4)} & {\bf (5)} & {\bf (6)}\\
 $\log (t_{k_{max}}/1\mathrm{~s})$ & $n_{max}$ & QCD model & $n_3$ & $n_4$ & Fig.\ \\
\hline
-9 	& 1.523	& CM,BM,LFM	& -0.0031	& 0.000078 & \ref{fig3}a\\ 	
\hline
-7 	& 1.599	& CM	& -0.0013	& -0.00030	& \ref{fig3}b\\ 
-7 	& 1.524	& BM	& -0.0021	& -0.00014	& \ref{fig3}b\\ 
-7 	& 1.593	& LFM	& -0.0014	& -0.00029	& \ref{fig3}b\\ 
\hline
-5 	& 1.686	& CM	& 0.0023	& -0.0012	& \ref{fig3}c\\
-5 	& 1.530	& BM	& -0.00030	& -0.00062 & \ref{fig3}c\\ 
-5 	& 1.587	& LFM	& 0.00064	& -0.00082 & \ref{fig3}c\\ 
\hline
{\bf -3} 	& {\bf 1.803}	& {\bf CM}	& {\bf 0.0099}	& {\bf -0.0033} & {\bf \ref{fig1},\ref{fig3}d} \\
-3 	& 1.732	& BM	& 0.0081	& -0.0028 & \ref{fig3}d\\ 
-3 	& 1.752	& LFM	& 0.0086	& -0.0030 & \ref{fig3}d\\ 
\hline
-1 	& 1.920	& CM,BM,LFM	& 0.026		& -0.0084 & \ref{fig3}e\\ 	
\hline
\hline
\end{tabular}
\end{table*}

\subsection{The mass spectrum of stellar mass PBHs}
\label{sec:5}

We can interpret the curve $n_{PBH}(t_{0},t_{k})$ (equation~\ref{mass-spectrum}) as a mass spectrum: the \emph{PBH mass spectrum}. For each of the cases shown in Figure~\ref{fig3} we divided the curve $n_{PBH}(t_{0},t_{k})$ into different portions, each corresponding to an order of magnitude, and integrated these in order to obtain the number density of PBHs of a given mass (equation \ref{n_t1_t2}). We have assumed, as a first approach, that PBHs formed at a particular epoch are uniformly distributed throughout the Universe. 

\begin{figure}
\centering
\includegraphics[width=70mm]{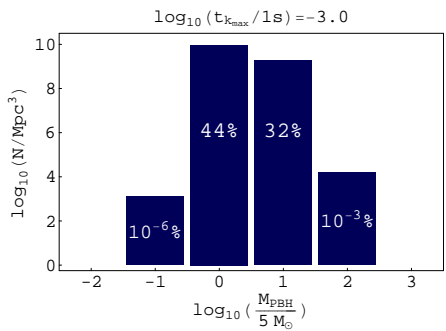} \\
\caption[]{The PBH mass mass spectrum when $t_{k_{max}}=10^{-3}\mathrm{~s}$ and $n_{max}=1.803$ (CM) -- cf. Figures \ref{fig1} and \ref{fig3}d. The values in percentage represent the contribution to the fraction of CDM in each case. See text and Table~\ref{table4} for more details.
\label{fig4}}
\end{figure}

The PBH mass spectrum of our best result is shown in Figure~\ref{fig4}  ($t_{k_{max}}=10^{-3}\mathrm{~s}$, $n_{max}=1.803$, QCD/CM -- see also Figures~\ref{fig1} and \ref{fig3}d),
 arising as a consequence of the plateau formed by the proximity of the radiation and QCD peaks (cf. Figure~\ref{fig3}d). 
From equation~(\ref{omega_t0}) we get, for this case, $\Omega_{PBH}\approx 0.197$. So, from Table~1, we get:

\begin{equation}
\label{Omega_76}
\frac{\Omega_{PBH}}{\Omega_{CDM}} = \frac{0.197}{0.258} = 0.763 \:.
\end{equation}
Thus, about 76\% of all CDM might be constituted by PBHs,
$44\%$ in the form of 5 M$_{\odot}$ and $32\%$ in the form of 50 M$_{\odot}$ ones.
Our second best result gives a 12\% contribution ($t_{k_{max}}=10^{-5}\mathrm{~s}$, $n_{max}=1.686$, again QCD/CM -- see Figure~\ref{fig3}c).
We have, thus, compiled all cases exemplified in this paper (Figures~2 and~3) in Table~\ref{table4},  focusing on results on the extended stellar mass range (0.05--500 M$_{\odot}$), which surely includes all stellar mass PBHs. 

\begin{table*}
\caption[]{The PBH stellar mass spectrum (0.05--500 M$_{\odot}$) according to the three QCD models (CM: Crossover Model; BM: Bag Model; LFM: Lattice Fit Model) for the relevant examples of Figure~\ref{fig3} (between curved brackets, the $\log (t_{k_{max}}$/1 s) is shown) that give results in that mass range in the most favourable situations (solid lines in Figure~\ref{fig2}), from where the $n_{max}$ values come from. The values shown in percentage 
between square brackets are the $>10^{-5}$~\% contribution to the fraction of CDM, in each case. 

\label{table4}
}
\center
\resizebox{\textwidth}{!}{
\begin{tabular}{cccccccccccc}
\hline \hline
M$_{\odot}$ & \multicolumn{11}{c}{N/Mpc$^3$}  \\
\cline{2-12}\\
 & $10^{-7}$ & $0.1$  & $1$  & $10$ & $100$  & $10^{3}$  & $10^{4}$  & $10^{8}$  & $10^{9}$  & $10^{10}$  & $10^{11}$   \\ \hline
0.05 & & & & & LFM(-7) & & CM(-7) & & & & CM(-5) [9\%]   \\ \hline
 & & & & & &  & & & BM(-7) [0.4\%] & CM(-5) [3\%] &    \\ 
 & & & & & & CM(-3)  & & & LFM(-7) [0.6\%] & BM(-3) [2\%] &    \\ 
0.5 & & & & & & BM(-9)  & & & BM(-5) [0.6\%] & LFM(-3) [2\%] &   \\ 
 & & & & & &  & & & LFM(-5) [0.7\%] &  &    \\ \hline
5 & & &BM(-3) & & &LFM(-3) & & CM(-5) [0.3\%] & &CM(-3) [44\%] &   \\ \hline
50 & BM(-3) & LFM(-3) & & & & & & &CM(-3) [32\%] & &   \\ \hline
 & & & & CM(-1) & & & & & & &   \\ 
500 & & & & BM(-1) & & & CM(-3) & & & &   \\ 
 & & & &  LFM(-1)& & & [0.001\%] & & & &   \\ \hline
\end{tabular}}
\end{table*}
\normalsize

\section{Discussion}
\label{sec:discussion}

The sources of many of the recently detected  gravitational waves by LIGO are likely BH binaries with masses within the range 18--85~M$_{\odot}$, suggesting an important population of binary BHs with stellar masses. However, it is not certain that these BHs  result from the final stages of stellar evolution. Instead, it is quite plausible  that these binaries are primordial in origin. So, this paper  explored scenarios for the formation of stellar mass PBHs (0.05--500~M$_{\odot}$). Although PBHs have not yet been observed directly in the Universe \citep[nevertheless, see, e.g.,][for interesting possibilities]{2014MNRAS.441.2878S} there are several observational constraints on the maximum number of PBHs of a given mass that could, eventually, have been formed at a given epoch.

PBHs can be formed from the collapse of overdense regions in the early Universe, provided that the amplitude of the density fluctuation is greater than some critical threshold $\delta_{c}$. During the QCD phase transition (when M$_{H}\sim$~0.5~M$_{\odot}$) the value of $\delta_{c}$ experiences a reduction which further favors PBH formation. We have, thus, studied three different models for the QCD phase transition (CM, BM, and LFM) using a running-tilt power-law spectrum for the primordial density fluctuations. We  selected five representative cases ($t_{k_{max}} = 10 ^{-9}, 10 ^ {-7}, 10 ^{-5}, 10 ^ {-3}$, and 10$^ {-1}\mathrm{~s}$), covering the full 0.05--500~M$_{\odot}$ range (corresponding to the range $1.4\leq n_{max}\leq 2.0$), with $t_{k_{max}}$ the instant when the fluctuation crosses the horizon and $n_{max}$ the amplitude of the spectral index at that instant.

There are about $2\times10^{12}$ galaxies in the Universe, with comoving densities of 0.1--1~Mpc$^{-3}$ and typical masses of 10$^{10}$~M$_{\odot}$ \citep[][]{2016ApJ...830...83C}. On a QCD/CM, in particular when $t_{k_{max}} = 10 ^{-3}\mathrm{~s}$,  we estimated  $10^9$--$10^{10}$~PBH/Mpc$^3$  (Table~3) with 5--50~M$_{\odot}$.
Therefore, from the comoving density of galaxies, one would
expect $\sim$~10$^{9-11}$~PBHs per galaxy.  


Although, at this stage, the actual model at the QCD epoch is not known, finding a monochromatic peak at $\sim$~0.5~M$_{\odot}$ will favour a BM or LFM model, while a broader mass spectrum
 (5--50M$_{\odot}$)
 will suggest a CM. Thus, if the latter applies,
 PBHs  are  excellent candidates for the observed gravitational wave cases, since their numbers could be as high as 76\% of the Universe CDM!
 

As regards future work we aim to consider the clustering of PBHs, in particular the formation of stellar mass PBH binaries that could account for the observed gravitational wave sources.

\end{document}